\documentstyle[aps,preprint,epsf]{revtex}

\tighten

\newcommand{\Tr}{{\rm Tr}}

\begin{document}

\preprint{UTCCP-P-20}


\title{Four-dimensional Simulation of the Hot Electroweak Phase
  Transition with the SU(2) Gauge-Higgs Model}

\author{Yasumichi Aoki}

\address{Center for Computational Physics,
  University of Tsukuba, Ibaraki 305, Japan}

\date{December 1996; revised May 1997}

\maketitle

\begin{abstract}

We study the finite-temperature electroweak phase transition of the minimal
standard model within the four-dimensional SU(2) gauge-Higgs model.
Monte Carlo simulations are performed  for intermediate values of the Higgs
boson mass in the range $50 \lesssim M_H \lesssim 100$GeV on a lattice with the
temporal size $N_t=2$.  The order of the transition is 
systematically examined using finite-size scaling methods. Behavior 
of the interface tension and the latent heat for an increasing Higgs 
boson mass is also investigated.
Our results suggest that the first-order transition
terminates around $M_H \sim 80$GeV.

\end{abstract}


\section{Introduction}

The possibility that the baryon number asymmetry of the 
universe was generated in the course of the electroweak phase
transition in the early universe has been much discussed. 
One of the conditions necessary for a realization of this possibility  
is the presence of out-of-equilibrium process during the transition.
If the transition is of first order with a sufficient strength,
super cooling could occur and the universe could have been driven out of
equilibrium.
This raises the question whether the electroweak phase transition undergoes a 
strong enough first-order transition for a realistic Higgs boson mass.

In the minimal standard model the essential features of the
finite-temperature electroweak phase transition are expected to be controlled
by the SU(2) gauge-Higgs sector.  Hence finite-temperature studies of the SU(2)
gauge-Higgs model represent a first step in understanding the electroweak
transition.  
In the past two approaches have been taken for a non-perturbative analysis
of the transition in this model by means of Monte Carlo simulations
(for reviews, see Refs.~\cite{RevLat95,RevLat96}).
One approach is to employ the original
four-dimensional model. The other is to
treat a three-dimensional model derived by a dimensional reduction of the
original model in time.
In the three-dimensional approach many studies have been made covering a 
wide range of the Higgs boson mass $35 \leq M_H \leq 180$GeV
\cite{Kajantie,Kajantie96,Leipzig3D,Karsch1,Karsch2}.  In particular,
it has recently been reported that the phase transition is absent for
$M_H \gtrsim 80$GeV\cite{Kajantie96,Karsch2}.
On the other hand, studies with the four-dimensional
model\cite{Leipzig92,Leipzig93,DESY94,DESY95,DESY96} have been less extensive,
and systematic analyses have been limited to relatively light Higgs boson
masses of $M_H\simeq 20-50$GeV\cite{DESY94,DESY95,DESY96}. 
Since the Higgs boson mass is experimentally bounded by
$M_H > 64$GeV\cite{Janot}, it is important to extend the study of the
four-dimensional model toward heavier Higgs boson masses. 

In this article, we report results of our simulation of the four-dimensional 
model for the Higgs boson mass in the range $50\lesssim M_H \lesssim 100$GeV.
Our aim is a systematic analysis of the strength of the transition,
which is known to be of strong first order for light Higgs boson,
as the Higgs boson mass $M_H$ increases.  For this purpose finite-size
scaling analyses of susceptibility and the Binder cumulant are carried out.
We also study latent heat and interface tension in order to examine the
strength of the transition in terms of physical quantities.   

In Sec.~\ref{sec:simulation} we describe some details of our simulation. Monte
Carlo time histories of observables are discussed in Sec.~\ref{sec:history}.
Results of finite-size scaling analyses are presented in Sec.~\ref{sec:fss}.
In Sec.~\ref{sec:d_e} and \ref{sec:sigma} we report measurement of latent heat
and interface tension.  Our results are summarized in Sec.~\ref{sec:summary}.
In Appendix we describe our procedure to estimate the Higgs boson mass $M_H$
and the critical temperature $T_c$ for our simulation points from published 
results.
%
\section{Simulation}
\label{sec:simulation}

We employ the standard lattice action given by 
\begin{equation}
\label{S_lat}
  S =  \sum_n \left\{
     \sum_{\mu > \nu} \frac{\beta}{2} \Tr U_{n,\mu \nu} 
    \mbox{} +  \sum_\mu \kappa \Tr
      ( \Phi_n^{\dag} U_{n, \mu} \Phi_{n+\hat{\mu}} )
    \mbox{} - \rho_n^2 - \lambda ( \rho_n^2 -1 )^2
    \right\},
\end{equation}
where the radial part $\rho_n$ is defined by the
decomposition of the complex $2\times2$ matrix Higgs field 
$\Phi_n = \rho_n \alpha_n , \; \rho_n \geq 0; \; \alpha_n \in$ SU(2).
All of our simulations are made for the temporal extent $N_t=2$.
We set the gauge coupling $\beta = 1/g^2 = 8$.  Simulations are made 
for  6 values of the scalar self-coupling $\lambda$, choosing  the hopping
parameter $\kappa$ in the vicinity of the transition point.
The parameter values of all of our runs are listed in Table \ref{tab:param}. 
Also listed in the table
are estimates of the zero-temperature Higgs boson mass
$M_H$ and the critical temperature
$T_c$
at the transition point of $N_t=2$ lattice.
These estimates are obtained by interpolating  available data for
$M_H$ and $T_c$ at
$\lambda = 0.0001$, $0.0003$, $0.0005$($M_H=47$GeV) \cite{DESY95,DESY96},
$0.0017235$($M_H=85$GeV) and $0.023705$ \cite{Leipzig92}.
The procedure of the interpolation is 
described
in Appendix.

For each value of $\lambda$ runs are made on an $N_s^3 \times 2$ lattice
with $N_s=8,12,16,24,32$,  and in addition with $N_s=40$ for $\lambda=0.001,
0.0017235, 0.003 ( M_H = 66, 85, 108$GeV ). Gauge and scalar fields are
updated with a combination of the heat  bath\cite{BunkLat95} and
overrelaxation\cite{OR}
algorithms in the ratio displayed in Table \ref{tab:ratio},
which has been reported to have the fastest
decorrelation time in terms of number of sweeps in Ref.~\cite{DESY95}.
We make $10^5$ iterations of the combined updates for each parameter point 
except for the run with $\lambda=0.0005, N_s=32, \kappa=0.128865$,
where we make $2\times10^5$ iterations.

Observables are measured at each iteration. Our analyses are mainly made in
terms of the quantities defined by 
\begin{eqnarray}
  \label{meas}
  L_s & \equiv &
    \frac{1}{N_s^3 N_t} \sum_n \frac{1}{3} \sum_{j=1}^{3}
    \left[
    \frac{1}{2} \Tr ( \Phi_n^{\dag} U_{n, j} \Phi_{n+\hat{j}} ) \right],\\
  \Lambda_s & \equiv &
    \frac{1}{N_s^3 N_t} \sum_n \frac{1}{3} \sum_{j=1}^{3}
    \left[
    \frac{1}{2} \Tr ( \alpha_n^{\dag} U_{n, j} \alpha_{n+\hat{j}} ) \right],\\
  \rho^2 & \equiv &
    \frac{1}{N_s^3 N_t} \sum_n \rho_n^2.
\end{eqnarray}
where $L_s$ is the spatial part of the scalar hopping term in the action, 
$\Lambda_s$ represents the angular part of $L_s$ and $\rho^2$
the squared Higgs condensate.

\section{Monte Carlo Time History}
\label{sec:history}

A qualitative test for a first-order phase transition is provided by
Monte Carlo time histories of observables and their histograms.
In Fig.~\ref{fig:history} we show the time history of $L_s$ and its
histogram for the largest volume for each $M_H$. For the lightest Higgs
boson $M_H = 47$GeV (Fig.~\ref{fig:history}(a)),
a clear flip-flop behavior is seen, which is reflected
in the double peak structure of the histogram of $L_s$.
If we increase the Higgs boson mass,
the flip-flop behavior becomes milder, indicating a weakening of the
first-order transition. It can be seen, however, up to
$M_H = 66$GeV (Fig.~\ref{fig:history}(d)).
A double peak structure of the histogram also persists up to
this value of $M_H$.  These observations provide qualitative indication for a
first-order transition up to $M_H = 66$GeV.

On the other hand, for the heavier Higgs boson of $M_H = 85$ and $108$GeV, we
cannot discern a flip-flop behavior nor a double peak structure in the
histogram.  We have analyzed the histogram varying the
hopping parameter $\kappa$ around the value of the simulation through the
reweighting technique\cite{reweighting}.
For $M_H=85$GeV (Fig.~\ref{fig:history}(e)), we found that the
histogram changes to a narrow flat peak  at $\kappa=0.129966$,
while only a sharp single peak structure has been  seen for
$M_H=108$GeV (Fig.~\ref{fig:history}(f)).
This suggests that a first-order transition is either absent, or
extremely weak at these Higgs boson masses.

\section{Finite-Size Scaling Analysis}
\label{sec:fss}

Let us consider the susceptibility of $\Lambda_s$ defined by 
\begin{equation}
  \label{chi}
    \chi_{\Lambda_s} \equiv V \left( \langle {\Lambda_s}^2 \rangle - \langle
{\Lambda_s}
      \rangle^2 \right),
\end{equation}
with  $V=N_s^3$ the spatial volume.  On a finite lattice this quantity
has a peak at the pseudo critical point $\tilde{\kappa_c}(V)$.  If the
transition is of first order, the maximum value of the peak should increase 
linearly in volume, and the pseudo critical point should converge
to the value at infinite volume linearly in $1/V$\cite{Barber}. On the
other hand, the maximum value should remain constant if there is no phase
transition.

We calculate the maximum value and the position of the peak with the standard
reweighting technique\cite{reweighting}.  Errors are estimated by a jackknife
analysis using $1000$-$20000$ sweeps as the bin size,
which is chosen from stability of
the magnitude of error as a function of bin size
and varies with coupling parameters and lattice sizes.

In Fig.~\ref{fig:sus} results for the maximum value of
the susceptibility of $\Lambda_s$ is shown as a function of the volume $V$.
For $47 \leq M_H \leq 66$GeV the increase of the maximum value is
consistent with a linear behavior, quantitatively supporting a first-order
transition for this region of Higgs boson mass. In contrast, we observe
a very flat volume dependence for $85\leq M_H \leq 108$GeV, albeit the
maximum value is increasing slowly in the range of volume used here.
Another useful indicator of a first-order transition is the Binder cumulant
defined by
\begin{equation}
  \label{Binder}
  B_{\Lambda_s} \equiv 1 - \frac{1}{3}
  \frac{\langle {\Lambda_s}^4 \rangle}{\langle {\Lambda_s}^2 \rangle^2}.
\end{equation}
For the case of a first-order phase transition the value of $B_{\Lambda_s}$ 
deviates from
$2/3$ in the infinite volume limit, while it converges to $2/3$ otherwise.

In Fig.~\ref{fig:Binder} we plot the valley depth of the Binder cumulant of
$\Lambda_s$ as a function of the inverse volume $1/V$ obtained by a reweighting
procedure similar to that for the susceptibility $\chi_{\Lambda_s}$.
Lines are linear fits to the largest three volumes for each $M_H$.
For $47 \leq M_H \leq 66$GeV the value extrapolated to the infinite
volume $B_{\Lambda_s}^{c}$ clearly deviates from $2/3$,
providing additional evidence for a
first-order transition. For $85 \leq M_H \leq 108$GeV the deviation 
decreases by an order of magnitude, although still finite within the error.
In Fig.~\ref{fig:Bind_ext} the deviation of $B_{\Lambda_s}^{c}$ from $2/3$ is
plotted as a function of the Higgs boson mass. It can be seen that the strength
of the first-order transition rapidly diminishes toward larger $M_H$.

Our results for the susceptibility and the Binder cumulant clearly show
that the transition is of first order for 
$47 \leq M_H \leq 66$GeV. It is also clear that the transition, if first
order, is a very weak one at
$M_H=85$GeV and 108GeV. It is  possible that the transition turns
into a crossover for this range of $M_H$.  Data for larger
volumes are needed, however, for a conclusive analysis on this point.

In Fig.~\ref{fig:k_c} we summarize our results for the pseudo critical point
$\tilde\kappa_c(V)$ obtained from the susceptibility and the Binder cumulant. 
In Table \ref{tab:k_c}
we summarize estimates of the critical hopping parameter in the infinite
volume obtained by a linear extrapolation of $\tilde\kappa_c(V)$ in $1/V$
using results of largest three volumes for each $\lambda$.
\section{Latent Heat}
\label{sec:d_e}

We calculate the latent heat with the equation \cite{d_e_d_rho},
\begin{equation}
  \label{CC}
  \Delta \epsilon \simeq - M_H^2 \kappa \Delta \langle \rho^2 \rangle.
\end{equation}
where $\Delta$ means the difference of $\langle \rho^2\rangle$ in the pure
symmetric phase and in the pure broken phase.  It has been
shown\cite{RevLat95,DESY96} that the latent heat
calculated with this equation is in good agreement with that obtained with
the energy operator.

In order to calculate the expectation value $\langle \rho^2\rangle$ in the
pure phases, we separate the ensemble of configurations generated in a run
into two sub-ensembles of pure phases.  The division is made by inspecting
the Monte Carlo time history of $\Lambda_s$.  
In order to avoid contaminations from transition stages,
iterations in those stages are removed following the method employed in
Ref.~\cite{qcdpax92}.

A source of systematic error in the procedure above arises from the fact
that the point of $\kappa$ at which the run is performed does not exactly
coincide with the pseudo critical point $\tilde{\kappa_c}$
which we estimate from the susceptibility $\chi_{\Lambda_s}$.
In order to treat this problem we make two independent runs,
both in the region of metastability,  such that
$\tilde{\kappa_c}$ is sandwiched between the hopping parameter of the
two runs. Then we interpolate results of the two runs for
$\langle \rho^2\rangle$
with a linear function of $\kappa$ to $\tilde{\kappa_c}$.
The error of the latent heat is calculated from those of
$\langle \rho^2\rangle$ and $\tilde{\kappa_c}$.

The result of our analysis is shown in Fig.~\ref{fig:d_e}.
At the lightest Higgs boson mass of $M_H=47$GeV our result is in good
agreement with that of Ref.~\cite{DESY95} obtained with the energy operator.
The latent heat decrease apparently linearly with an increasing Higgs boson
mass $M_H$.  A linear
extrapolation of our data is consistent with a vanishing of the latent heat at 
$M_H\sim 80$GeV.

\section{Interface Tension}
\label{sec:sigma}

The interface tension $\sigma$ provides another indicator of the strength
of the first-order transition. We calculate this quantity with the Binder's
histogram method \cite{Binder}. Let
$P_{max}$ and $P_{min}$ be the peak and valley height of the distribution of
$\Lambda_s$, reweighted such that the two peaks have an equal height. Define 
$\hat\sigma_V \equiv - ({N_t^2}/{2 N_s^2}) \ln ({P_{min}}/{P_{max}}) $. 
For  spatially cubic lattices used in our simulations, finite-size formula for
the true interface tension $\sigma$ is
given by \cite{Kanaya}
\begin{equation}
  \label{ift_corr}
  \hat\sigma_V (N_s, N_t) = \frac{\sigma}{T_c^3} - \frac{N_t^2}{N_s^2}
  \left[ c - \frac{1}{4} \ln N_s
  \right],
\end{equation}
where $c$ is a constant independent of $N_s$. 
Making a two parameter fit of
$\hat\sigma_V$ obtained for the largest three volumes for $M_H = 47, 52, 
57$GeV or two volumes for $M_H = 66$GeV we find $\sigma/T_c^3$
shown in Fig.~\ref{fig:ift}.
The interface tension rapidly decrease with an increasing Higgs boson mass
and seems to vanish around $M_H \sim 80$GeV as in the case of the latent heat.

\section{Summary}
\label{sec:summary}

Our finite-size scaling study establishes a first-order transition for
$47 \leq M_H \leq 66$GeV.
For larger Higgs boson masses a rapid weakening of the
transition makes it difficult to draw a definitive conclusion on the order
within the range
of lattice volumes employed in our simulation.  However, combining
finite-size data  with results for the latent heat and the interface tension,
our four-dimensional study suggests that the first-order transition terminates
around $M_H \sim 80$GeV in the $N_t=2$ SU(2) gauge-Higgs model. This is
consistent with the results of recent finite-size scaling studies carried out
in the dimensionally reduced three-dimensional model\cite{Kajantie96,Karsch2}.

\section*{Acknowledgements}
I would like to thank Akira Ukawa for useful discussions.
The numerical calculations were carried out on VPP/500
at Science Information Processing Center of University of Tsukuba
and at Center for Computational Physics at University of Tsukuba.

\appendix
\section{}

In Table \ref{tab:param} we listed the zero temperature Higgs boson mass 
$M_H$ and the critical temperature $T_c$ for each $\lambda$
at our critical hopping parameter $\kappa_c$ on the $N_t=2$ lattice.
Here we describe how we estimated those values.

In Table \ref{tab:scale} we summarize published results for $M_H$ and $T_c$ 
at five different values of $\lambda$ reported in 
Refs.~\cite{Leipzig92,DESY95,DESY96}.
Two values of $\lambda$ employed in our simulation 
($\lambda=0.0005,0.0017235$ ) are among those in Table ~\ref{tab:scale}. 
For these cases we adopt the published results for  $M_H$ and $T_c$.
For the other four values of $\lambda$ we estimate $M_H$ and $T_c$ by an 
interpolation of results in Table ~\ref{tab:scale}.

At the tree revel the mass ratio $M_H/M_W$
is proportional to $\sqrt{\lambda}/\kappa$ 
(the gauge coupling is fixed to $\beta=8$ in our simulations).
Plotting five values of $M_H$ in Table ~\ref{tab:scale} against 
$\sqrt{\lambda}/\kappa$ we find that they exhibit a smooth increase 
suitable for an interpolation.
A cubic polynomial of $\sqrt{\lambda}/\kappa$ yields a good fit with 
$\chi^2/\mbox{dof}=0.10$.  We then estimate the 
values of $M_H$ at $\lambda=0.000625, 0.00075, 0.001, 0.003$ from the 
value of the fitted curve at $\sqrt{\lambda}/\kappa_c$ 
where $\kappa_c$ is the critical point determined from the 
susceptibility of $\Lambda_s$ listed in Table \ref{tab:k_c}.
One can also make a fit with a quadratic polynomial 
($\chi^2/\mbox{dof} = 0.05$).
The difference of values of $M_H$ from the two fits may be used as a measure 
of the systematic error of our estimation procedure. However, the difference
is quite small ( $0.03-0.3\%$) compared with the statistical error
($2-6\%$), and we adopt the statistical error in our estimate.

The results for the critical temperature $T_c$ has a sharp bend for large 
$\sqrt{\lambda}/\kappa$.  Hence this variable is not suitable for 
interpolation of $T_c$.  We find, however, that   
$T_c$ as a function of $M_H$ is smoothly increasing, and a cubic polynomial fit
goes well ($\chi^2/\mbox{dof} = 0.5$).  We use the fitted curve and the values of 
$M_H$ estimated above to find the values of $T_c$.
Fit with a quadratic polynomial ($\chi^2/\mbox{dof} = 0.9$)
is used to estimate the systematic error, which
turns out to be comparable to the statistical one.
Those errors are combined into the final value of error shown in
Table \ref{tab:param}.


\begin{table}[p]
  \begin{center}
    \caption{Run parameters of simulation.
             For each parameter $10^5$ sweeps are made except for the run
             marked by $\dag$ ($2\times10^5$ sweeps).
             Data generated are used for analysis of susceptibility($\chi$),
             Binder cumulant($B$), latent heat($\Delta\epsilon$) and
             interface tension($\sigma$).
                For our procedure to estimate the Higgs boson mass $M_H$ and 
                the critical temperature $T_c$, see Appendix.
                }
    \label{tab:param}
    \leavevmode

\vspace{-6pt}
    \begin{tabular}{lccrrl}
    \mbox{\hspace{20pt}}$\lambda$ & $M_H$(GeV) & $T_c$(GeV) & $N_s$ &
     \mbox{\hspace{16pt}}$\kappa$\mbox{\hspace{16pt}} & use \\
    \hline
    0.0005    &  47(2) &  94(1)  &  8 & 0.128950 & $\chi$  \\
                            &&& 12 & 0.128900 & $\chi,B$  \\
                            &&& 16 & 0.128883 & $\chi,B,\sigma$  \\
                            &&& 24 & 0.128866 & $\chi,B,\sigma$  \\
                            &&& 32 & 0.128860 & $B$  \\
                            &&& 32 & 0.128862 & $\Delta\epsilon$\\
                     &&& $\dag$ 32 & 0.128865 & $\chi,\Delta\epsilon,\sigma$ \\
    \\
    0.000625  &  52(1) & 100(3)  &  8 & 0.129110 & $\chi$  \\
                            &&& 12 & 0.129036 & $\chi,B$  \\
                            &&& 16 & 0.129004 & $\chi,B,\sigma$  \\
                            &&& 24 & 0.128986 & $\chi,B,\sigma$  \\
                            &&& 32 & 0.128983 & $B,\Delta\epsilon$  \\
                            &&& 32 & 0.128987 & $\chi,\Delta\epsilon,\sigma$ \\
    \\
    0.00075   &  57(1) & 107(3)  &  8 & 0.129243 & $\chi$  \\
                            &&& 12 & 0.129158 & $\chi,B$  \\
                            &&& 16 & 0.129126 & $\chi,B,\sigma$  \\
                            &&& 24 & 0.129103 & $\chi,B,\sigma$  \\
                            &&& 32 & 0.129098 & $B$   \\
                            &&& 32 & 0.129102 & $\chi,\Delta\epsilon,\sigma$ \\
                            &&& 32 & 0.129106 & $\Delta\epsilon$\\
    \end{tabular}
  \end{center}
\end{table}

\addtocounter{table}{-1}
\begin{table}[p]
  \begin{center}
  \caption{{({\it Continued}\/)}}
    \leavevmode

    \begin{tabular}{lccrrl}
    \mbox{\hspace{20pt}}$\lambda$ & $M_H$(GeV) & $T_c$(GeV) & $N_s$ &
     \mbox{\hspace{16pt}}$\kappa$\mbox{\hspace{16pt}} & use \\
    \hline
    0.001     &  66(2) & 119(3)  &  8 & 0.129476 & $\chi$   \\
                            &&& 12 & 0.129407 & $\chi,B$  \\
                            &&& 16 & 0.129349 & $\chi,B$  \\
                            &&& 24 & 0.129330 & $\chi,B$  \\
                            &&& 32 & 0.129328 & $\chi,B,\sigma$  \\
                            &&& 40 & 0.129327 & $\chi,B,\Delta\epsilon,\sigma$ \\
                            &&& 40 & 0.129330 & $\Delta\epsilon$ \\
    \\
    0.0017235 &  85(3) & 139(2) &  8 & 0.130200 & $\chi$  \\
                            &&& 12 & 0.130080 & $\chi,B$  \\
                            &&& 16 & 0.130026 & $\chi,B$  \\
                            &&& 24 & 0.129980 & $\chi,B$  \\
                            &&& 32 & 0.129966 & $\chi,B$  \\
                            &&& 40 & 0.129968 & $\chi,B$  \\
    \\
    0.003     & 108(6) & 157(10) &  8 & 0.131350 & $\chi$  \\
                            &&& 12 & 0.131050 & $B$  \\
                            &&& 12 & 0.131200 & $\chi$  \\
                            &&& 16 & 0.131111 & $\chi,B$  \\
                            &&& 24 & 0.131065 & $\chi,B$  \\
                            &&& 32 & 0.131054 & $\chi,B$  \\
                            &&& 40 & 0.131042 & $\chi,B$  \\
    \end{tabular}
  \end{center}
\end{table}

\begin{table}[p]
  \begin{center}
    \caption{Number of iteration of each updating step in one sweep.}
    \label{tab:ratio}
    \leavevmode

    \begin{tabular}{@{\hspace{12pt}}cc@{\hspace{12pt}}|@{\hspace{12pt}}ccc}
      \multicolumn{2}{c|@{\hspace{12pt}}}{heat bath} &
      \multicolumn{3}{c}{overrelaxation} \\
      $U$ & $\Phi$ & $U$ & $\alpha$ & $\rho$\\
      \hline
      1 & 4 & 3 & 3 & 1 \\
    \end{tabular}
  \end{center}
\end{table}

\begin{table}[p]
  \begin{center}
    \caption{Critical hopping parameter $\kappa_c$ for each $\lambda$.}
    \label{tab:k_c}
    \leavevmode

    \begin{tabular}{c|cccccc}
      $\lambda$ & 0.0005 & 0.000625 & 0.00075 & 0.001 & 0.0017235 & 0.003 \\
      \hline
      \hline
      $\kappa_c$ ($\chi_{\Lambda_s}$) & 0.1288607(23) & 0.1289812(11)
        & 0.1290981(14) & 0.1293262(11) & 0.1299633(29) & 0.1310360(40)\\
      $\kappa_c$ ($B_{\Lambda_s}$) & 0.1288584(13) & 0.1289822(25)
        & 0.1290974(14) & 0.1293266(11) & 0.1299622(27) & 0.1310318(34)\\
    \end{tabular}
  \end{center}
\end{table}

\begin{table}[p]
  \begin{center}
    \caption{Summary of zero temperature Higgs boson mass $M_H$ and critical
      temperature $T_c$ at the transition point of $N_t=2$ lattice.}
    \label{tab:scale}
    \leavevmode
    \begin{tabular}{llcrc}
      \multicolumn{1}{c}{$\lambda$} & \multicolumn{1}{c}{$\kappa$}
      & \multicolumn{1}{c}{$M_H$(GeV)} & \multicolumn{1}{c}{$T_c$(GeV)}
      & reference\\
      \hline
      0.0001    & 0.1283  &  18(1)  &  38(1) & \cite{DESY95}\\
      0.0003    & 0.12865 &  35(1)  &  71(1) & \cite{DESY96}\\
      0.0005    & 0.12885 &  47(2)  &  94(1) & \cite{DESY95}\\
      0.0017235 & 0.13    &  85(3)  & 139(2) & \cite{Leipzig92}\\
      0.023705  & 0.145   & 196(13) & 185(4) & \cite{Leipzig92}\\
    \end{tabular}
  \end{center}
\end{table}
%
%


%
\begin{figure}[p]
\vspace*{1cm}
  \begin{center}
    \leavevmode
    \epsfxsize=15cm \epsfbox{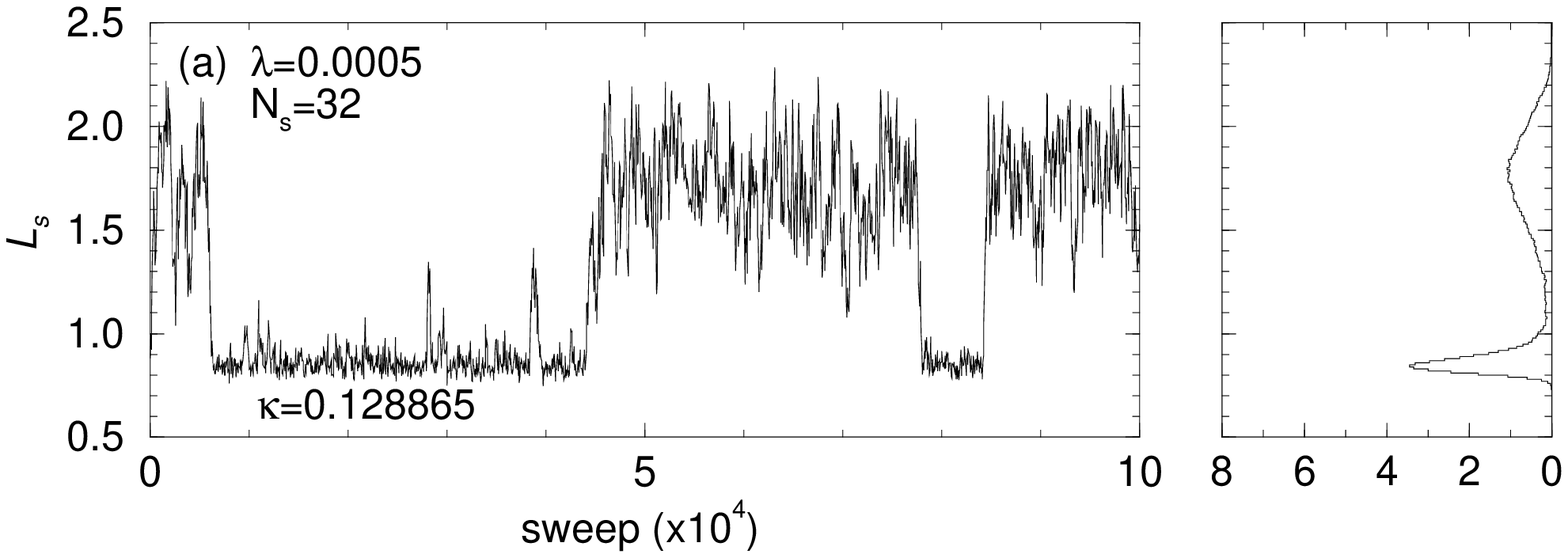}
  \end{center}
  \begin{center}
    \leavevmode
    \epsfxsize=15cm \epsfbox{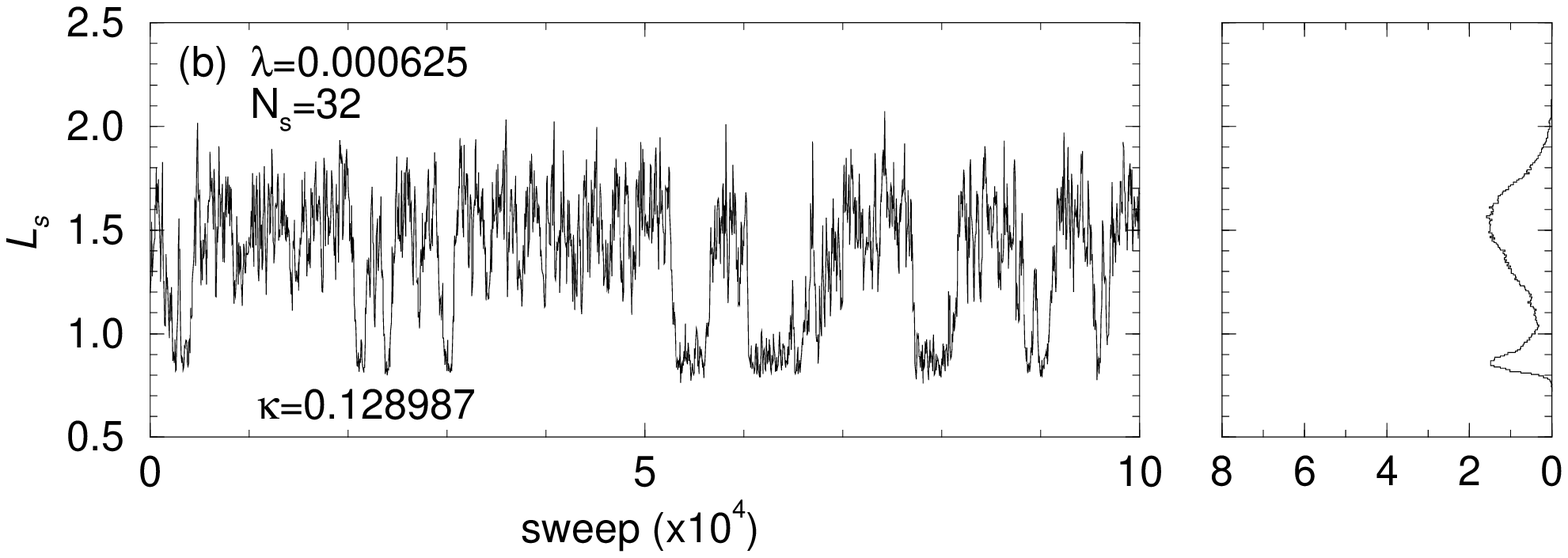}
  \end{center}
  \begin{center}
    \leavevmode
    \epsfxsize=15cm \epsfbox{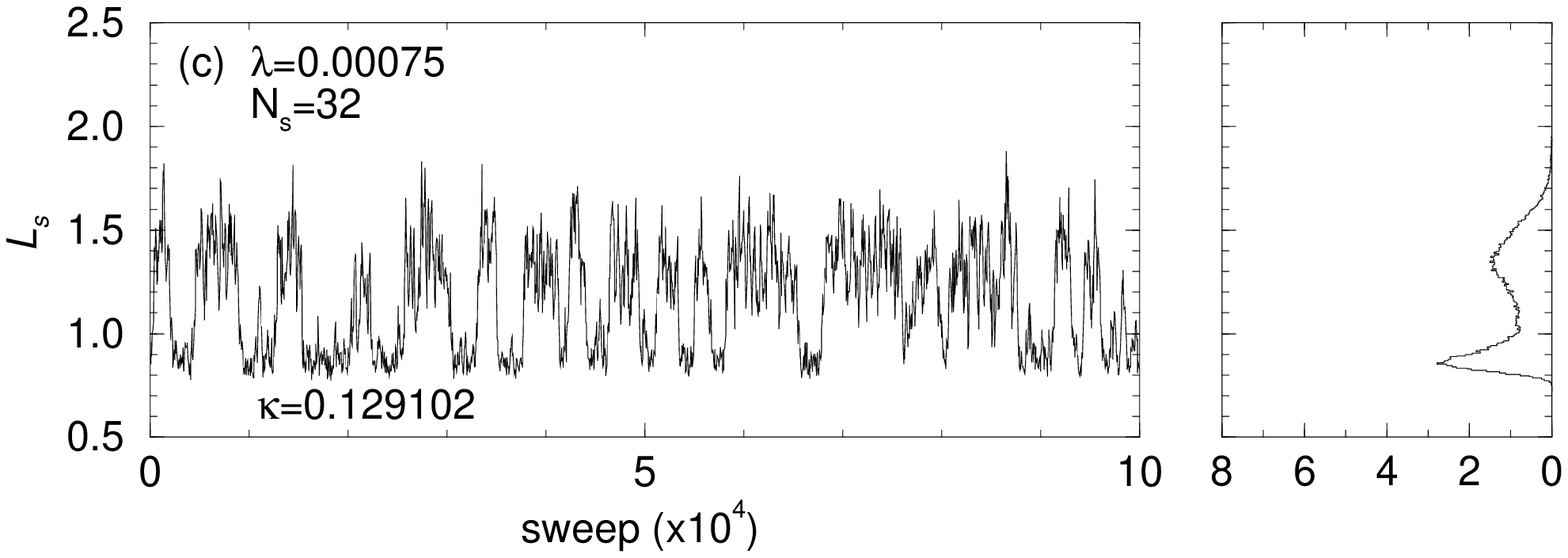}
  \end{center}
  \caption{Monte Carlo time history of the spatial hopping term $L_s$
    and its histogram.
    Data at every 50th sweep are displayed for history.
    Histograms are made with full ensemble.
    (a) $M_H=47$GeV, (b) $M_H=52$GeV and (c) $M_H=57$GeV.}
  \label{fig:history}
\end{figure}

\addtocounter{figure}{-1}
\begin{figure}[p]
\vspace*{1.5cm}
  \begin{center}
    \leavevmode
    \epsfxsize=15cm \epsfbox{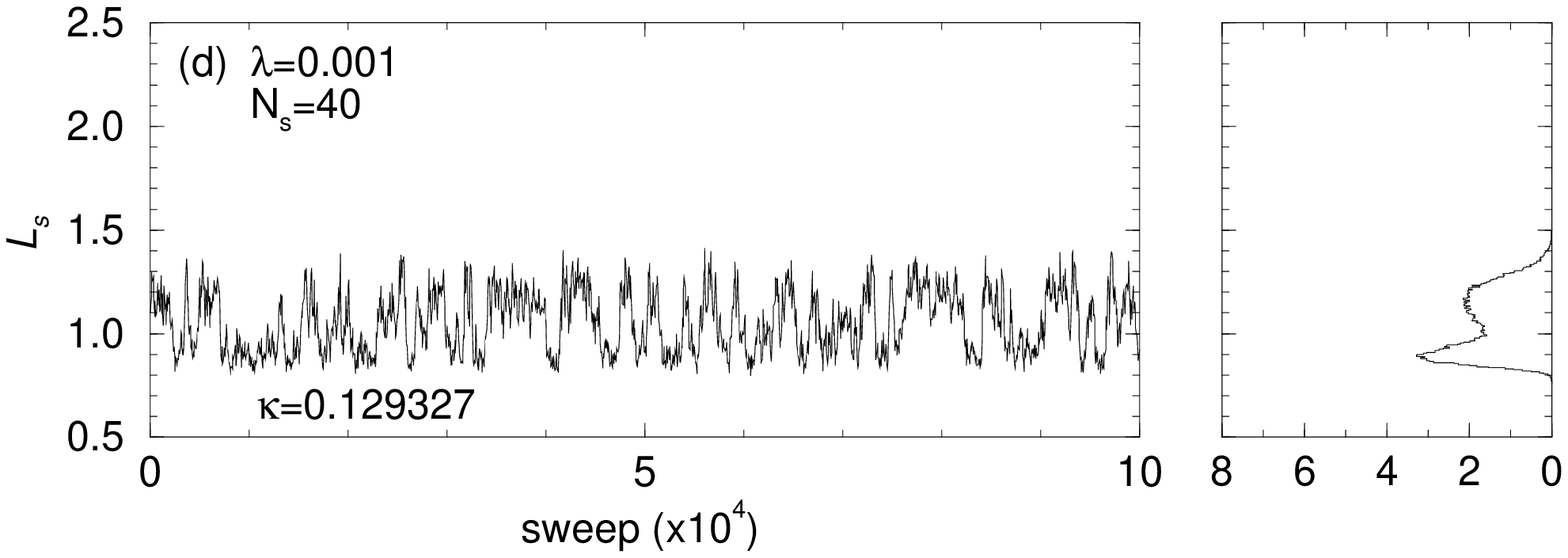}
  \end{center}
  \begin{center}
    \leavevmode
    \epsfxsize=15cm \epsfbox{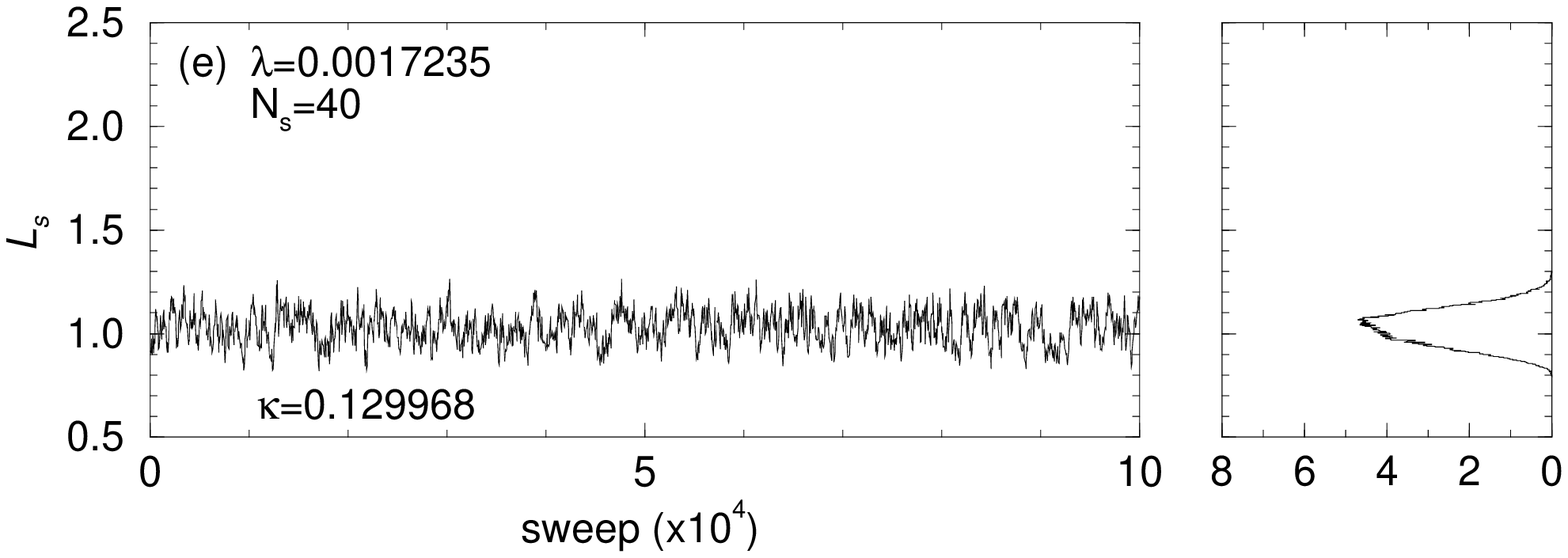}
  \end{center}
  \begin{center}
    \leavevmode
    \epsfxsize=15cm \epsfbox{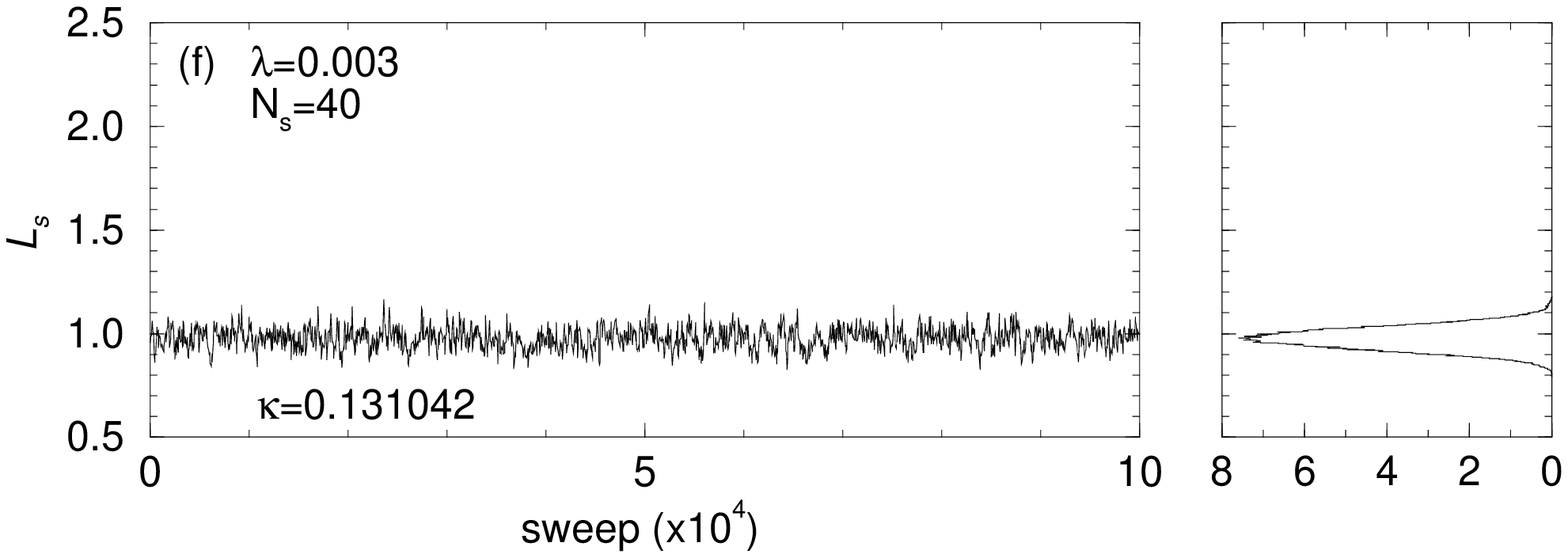}
  \end{center}
  \caption{{({\it Continued}\/)}
    (d) $M_H=66$GeV, (e) $M_H=85$GeV and (f) $M_H=108$GeV.}
\end{figure}

\begin{figure}[p]
\vspace*{1cm}
  \begin{center}
    \leavevmode
    \epsfxsize=11cm \epsfbox{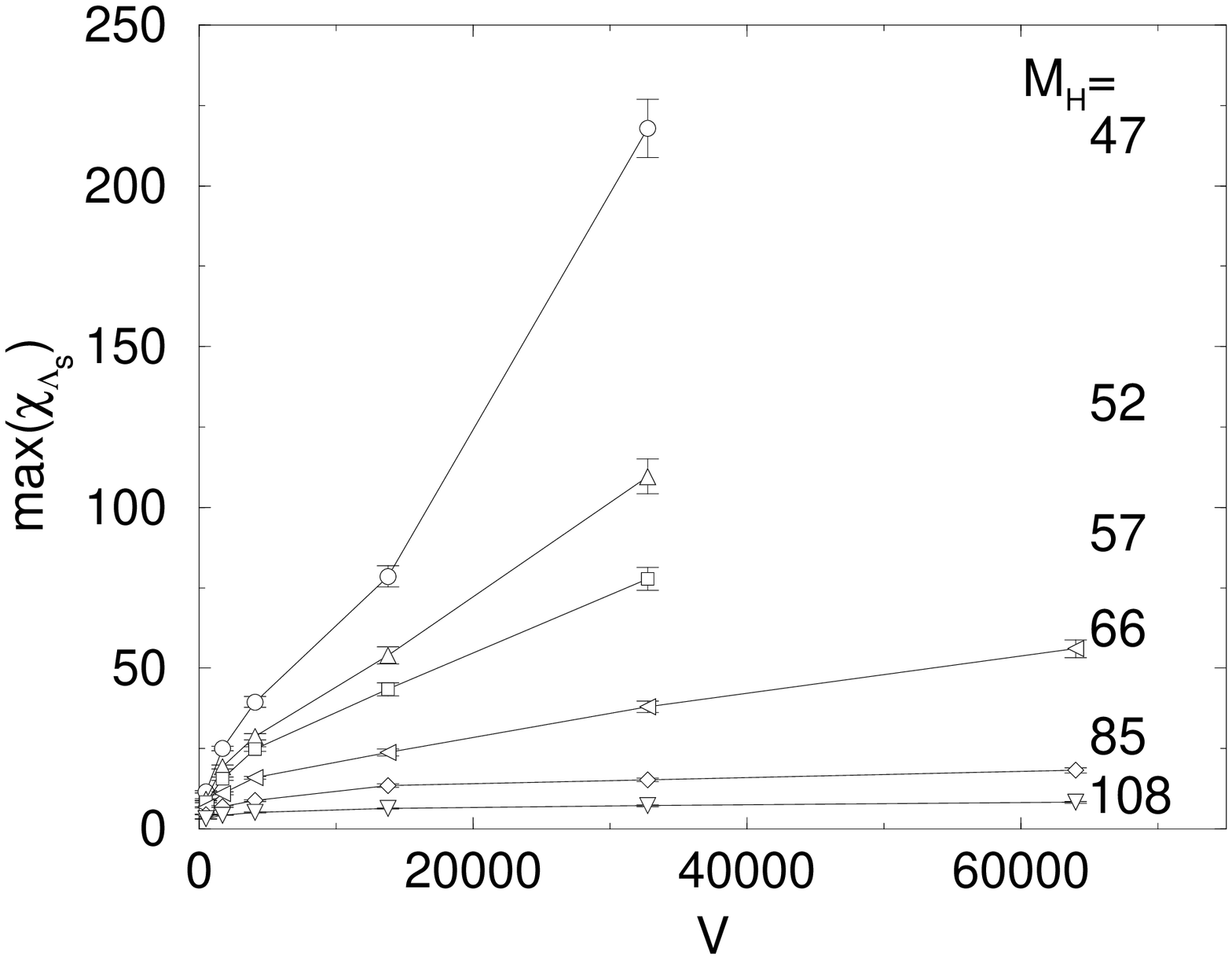}
  \end{center}
  \caption{Maximum height of the susceptibility of $\Lambda_s$ as a function
    of the volume. Lines are guides for eyes.}
  \label{fig:sus}
\end{figure}

\begin{figure}[p]
  \begin{center}
    \leavevmode
    \epsfxsize=11cm \epsfbox{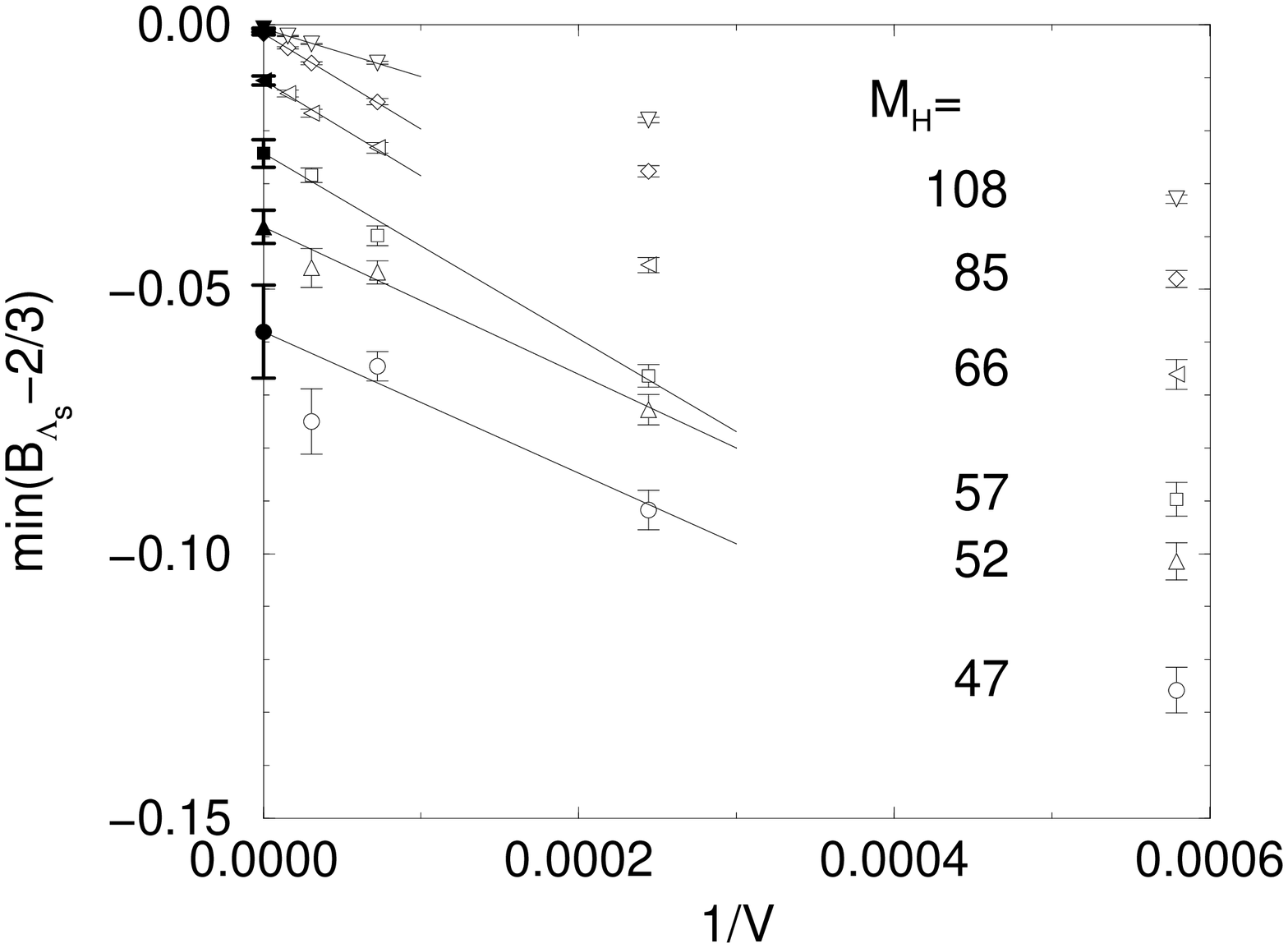}
  \end{center}
  \caption{Valley depth of Binder cumulant of $\Lambda_s$ as a function
    of the inverse volume. Filled symbols show the infinite volume limit.
    }
  \label{fig:Binder}
\end{figure}

\begin{figure}[p]
\vspace*{3cm}
  \begin{center}
    \leavevmode
    \epsfxsize=12cm \epsfbox{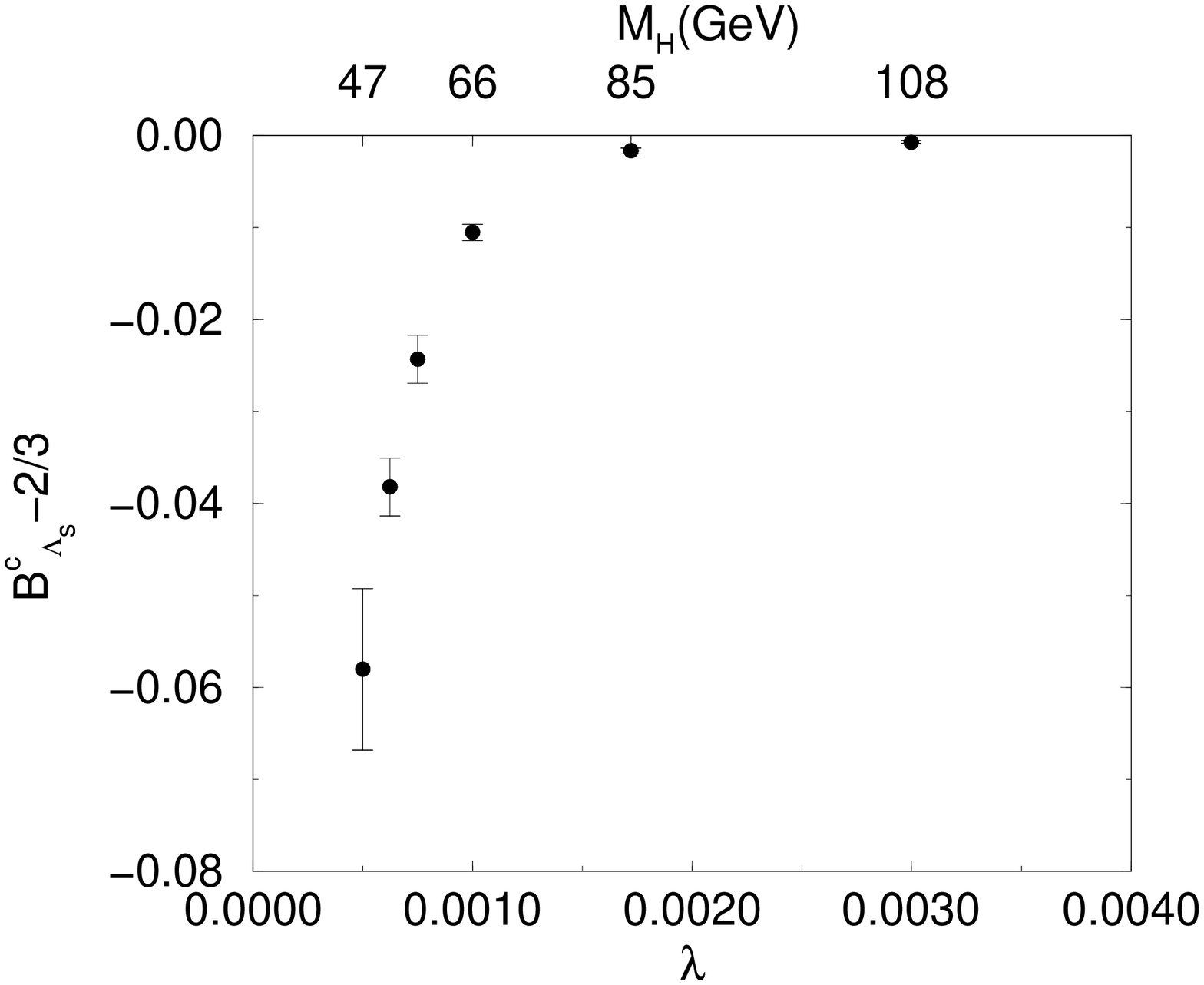}
  \end{center}
  \caption{Valley depth of $B_{\Lambda_s}$ at infinite volume
    as a function of Higgs boson mass.
    }
  \label{fig:Bind_ext}
\end{figure}

\begin{figure}[p]
\vspace*{3cm}
  \begin{center}
    \leavevmode
    \epsfxsize=14cm \epsfbox{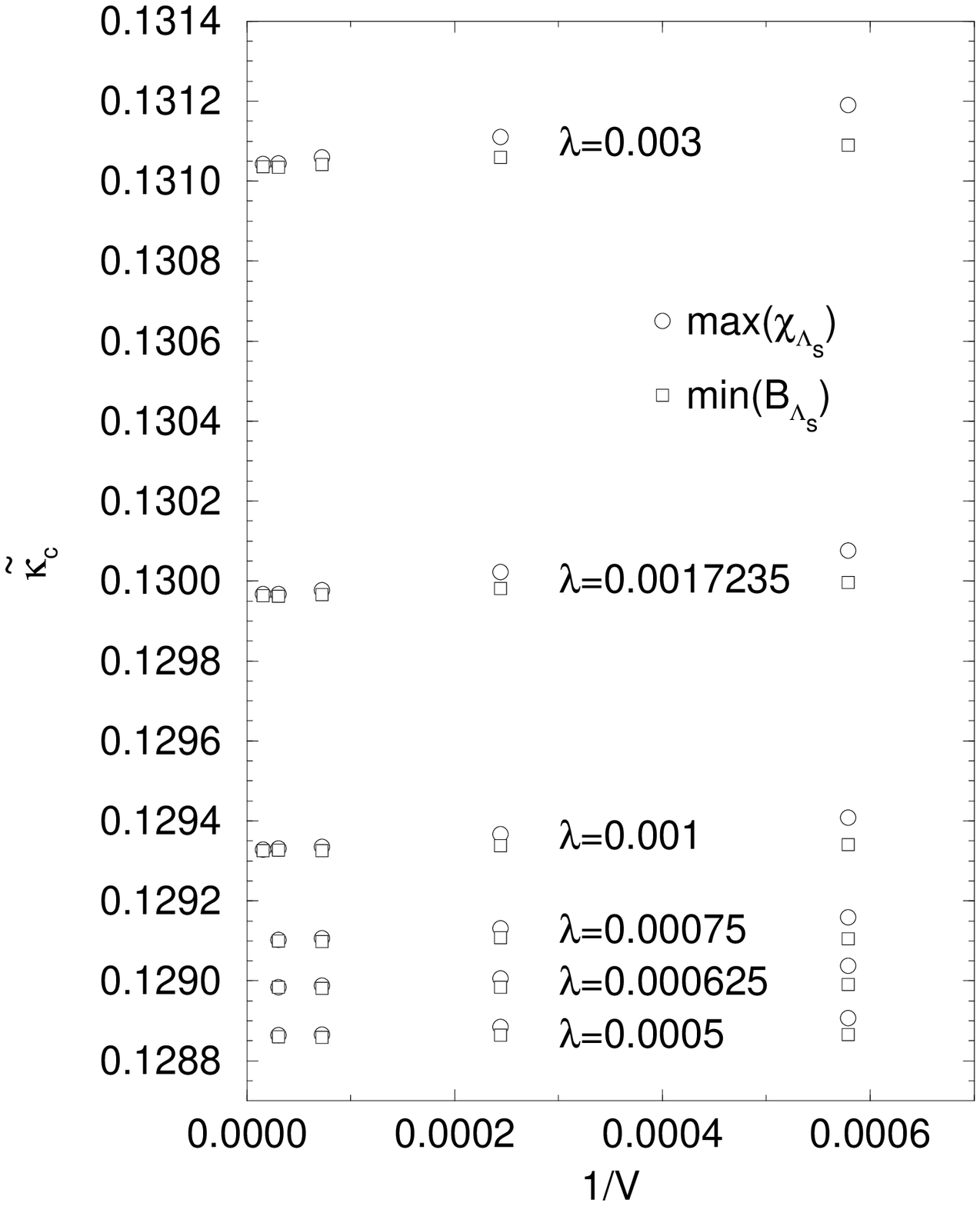}
  \end{center}
  \caption{Pseudo critical hopping parameter $\tilde{\kappa_c}$ defined by the
    maximum point of $\chi_{\Lambda_s}$ or
    the minimum point of $B_{\Lambda_s}$.
    }
  \label{fig:k_c}
\end{figure}

\begin{figure}[p]
\vspace*{3cm}
  \begin{center}
    \leavevmode
    \epsfxsize=12cm \epsfbox{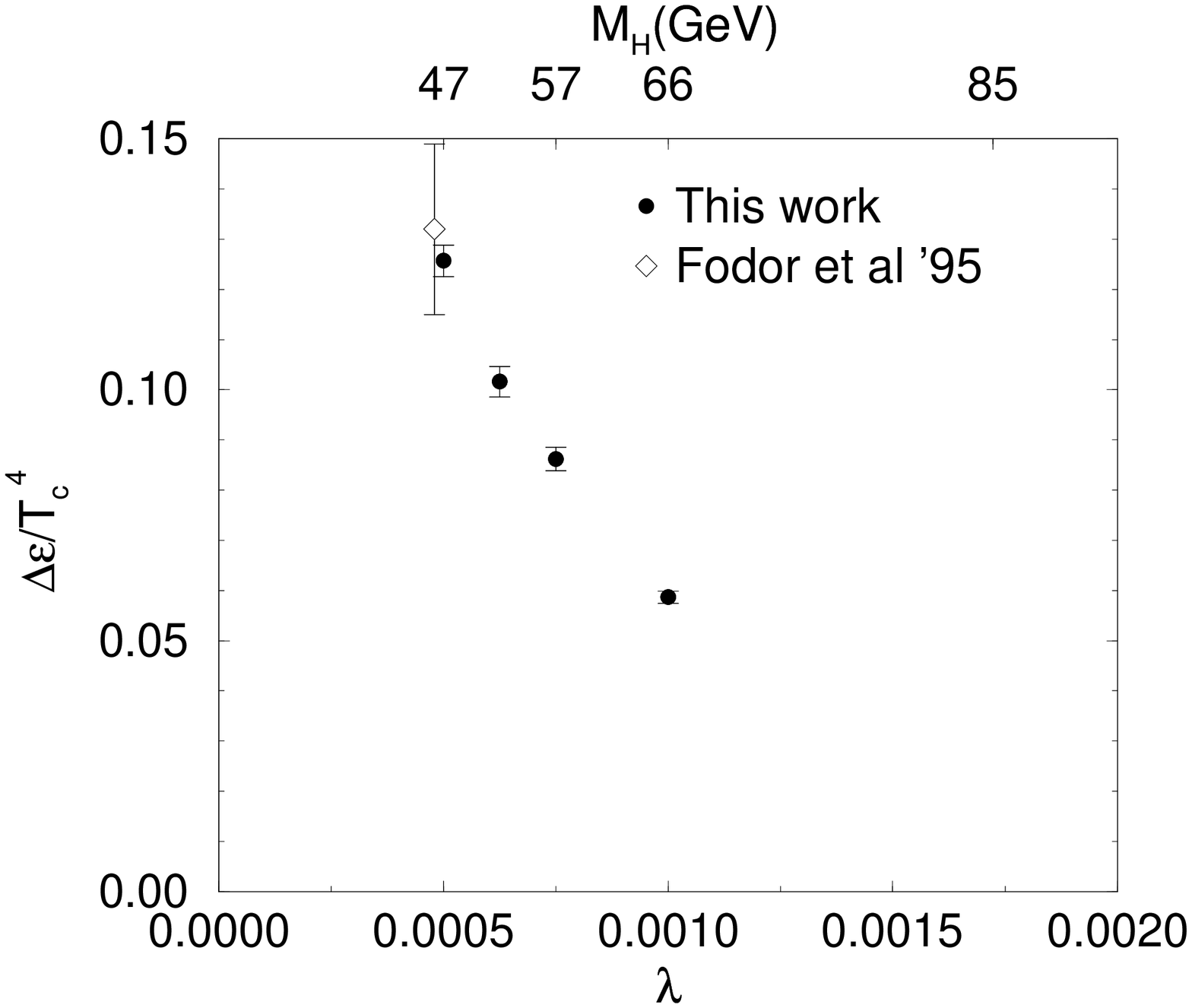}
  \end{center}
  \caption{Latent heat as a function of Higgs boson mass.
    Open symbol is from
    Ref.~\protect\cite{DESY95}
    and is shifted slightly left for visualization.}
  \label{fig:d_e}
\end{figure}

\begin{figure}[p]
  \begin{center}
    \leavevmode
    \epsfxsize=12cm \epsfbox{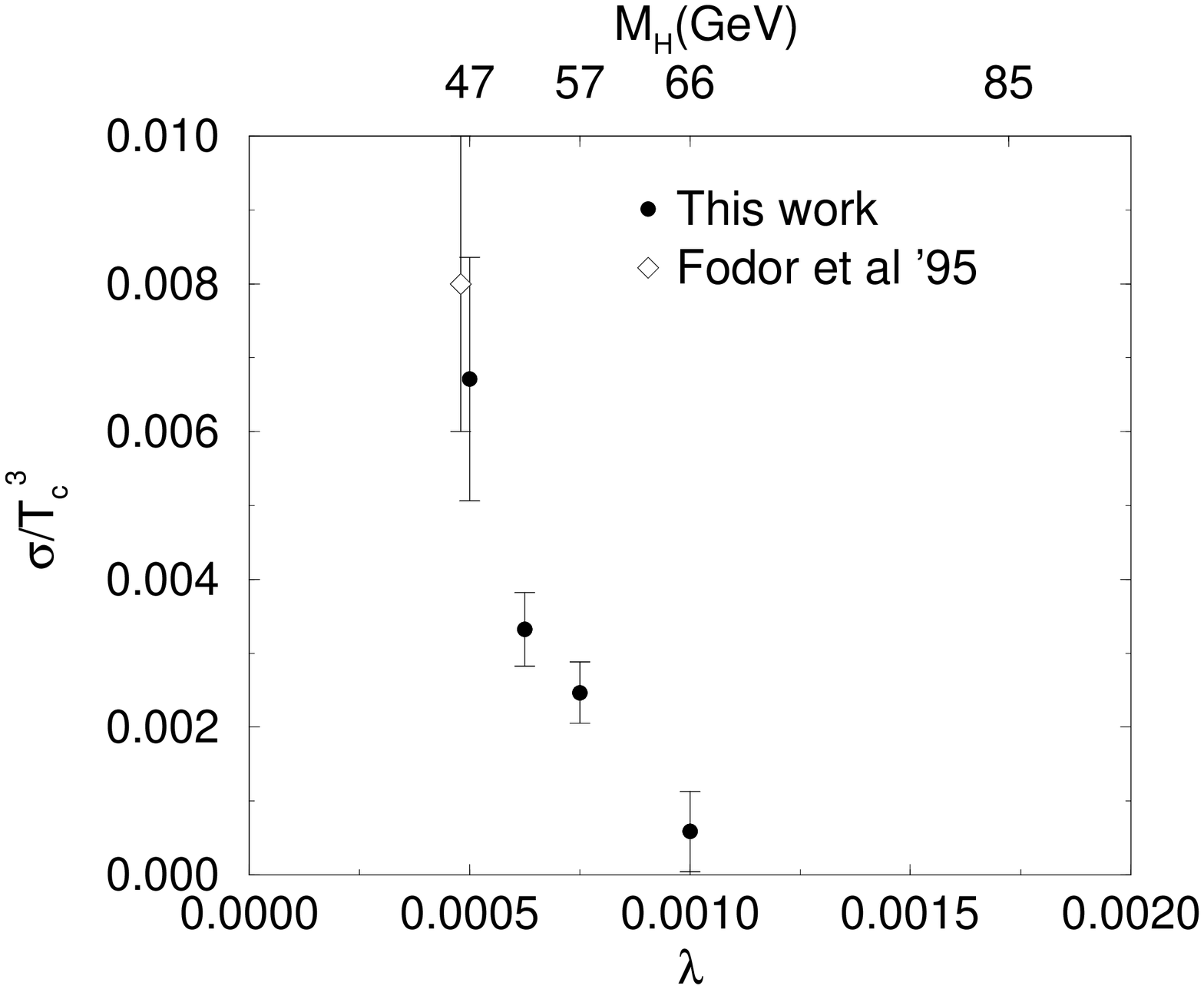}
  \end{center}
  \caption{Interface tension as a function of Higgs boson mass.
    Legends are the same as in
    Fig.~\protect\ref{fig:d_e}.}
  \label{fig:ift}
\end{figure}

\end{document}